\documentclass[aip,twocolumn,letterpaper]{revtex4}

\usepackage{fullpage}

\usepackage{graphics}
\usepackage{epsfig}

\usepackage[margin=1.0in]{geometry}

\begin{document}
\title{Magnetic field evolution and reconnection in low resistivity plasmas}
\author{Allen H Boozer}
\affiliation{Columbia University, New York, NY  10027\\ ahb17@columbia.edu}

\begin{abstract}

The mathematics and physics of each of the three aspects of magnetic field evolution---topology, energy, and helicity---is remarkably simple and clear.  When the resistivity $\eta$ is small compared to an imposed evolution, $a/v$, timescale, which means $R_m\equiv\mu_0va/\eta>>1$, magnetic field line chaos dominates the evolution of field-line topology in three-dimensional systems.  Chaos has no direct role in the dissipation of energy.  A large current density, $j_\eta\equiv vB/\eta$, is required for energy dissipation to be on a comparable time scale to the topological evolution.  Nevertheless, chaos plus Alfv\'en wave damping  explain why both timescales tend to be approximately an order of magnitude longer than the evolution timescale $a/v$.   Magnetic helicity is injected onto tubes of field lines when boundary flows have vorticity.  Chaos can spread but not destroy magnetic helicity.  Resistivity has a negligible effect on helicity accumulation when $R_m>>1$.  Helicity accumulates within a tube of field lines until the tube erupts and moves far from its original location.

%The recognized challenges of a driven evolution of a magnetic field are addressed in its three aspects:  field line topology, magnetic energy, and magnetic helicity.   Magnetic field lines can go from a simple smooth form to having large and broadly-spread changes in their connections on a timescale that is approximately a factor of ten longer than the ideal evolution time when and only when the magnetic field lines become chaotic.  Footpoint motions transfer magnetic energy to coronal loops, which can only be balanced by highly localized current densities $j\approx vB/\eta$, where $v$ is the footpoint velocity and $\eta$ is the plasma resistivity.  These current densities are consistent with those required to produce the solar corona with the observed height of the transition region through the Dreicer electron runaway effect.   A small resistivity cannot balance the magnetic helicity input by footpoint motion and leads to the eruption of coronal loops.

\end{abstract}

\date{\today} 
\maketitle

%\tableofcontents

\section{Introduction}

According to Wikipedia: \emph{Magnetic reconnection is a physical process occurring in highly conducting plasmas in which the magnetic topology is rearranged and magnetic energy is converted to kinetic energy, thermal energy, and particle acceleration.}  

By definition, a highly conducting plasma has a magnetic Reynolds number,
\begin{equation}
R_m\equiv \frac{\mu_0 v a}{\eta},
\end{equation}
that is far larger than unity; $v$ is a typical plasma flow speed, $a$ a typical spatial scale across the magnetic field, and $\eta/\mu_0$ is the resistive diffusion coefficient.  In problems of interest $R_m$ can be between $10^4$ and $10^{14}$.

Magnetic topology rearrangement and magnetic energy conversion are distinct physical concepts, but each concept has been used to define magnetic reconnection.  The classical definition was in a 1956 paper of Parker and Krook \cite{Parker-Krook:1956}: \emph{severing and reconnection of lines of force}.  In the space sciences, the emphasis has focused on energy conversion.  In 2020, Hesse and Cassak stated \cite{Hesse-Cassak}: \emph{Magnetic reconnection converts, often explosively, stored magnetic energy to particle energy in space and in the laboratory.}   

A practical understanding of magnetic field evolution when the magnetic Reynolds number, $R_m$, is large requires understanding the evolution of three distinct concepts: (1) magnetic topology, (2) magnetic energy, and (3) magnetic helicity.  This paper was written to disentangle the evolution of these three concepts, which are confusingly entangled in discussions of magnetic reconnection.

The simple Ohm's Law, $\vec{E}+\vec{v}\times\vec{B}=\eta\vec{j}$ illustrates the paradox of reconnection theory when the evolution of magnetic topology, energy, and helicity are entangled.   The combination of Ohm's Law, Faraday's Law, and Ampere's Law yields two equations
\begin{eqnarray}
\frac{\partial\vec{B}}{\partial t}=\vec{\nabla}\times\big(\vec{v}\times\vec{B}-\eta\vec{j}\big) \label{B-evol} \mbox{   and  }\\
\frac{\partial\vec{B}}{\partial t} -\vec{\nabla}\times(\vec{v}\times\vec{B})=\frac{\eta}{\mu_0}\nabla^2\vec{B} \label{B-adv}.
\end{eqnarray}

Equation (\ref{B-evol}) implies that the resistivity $\eta$ competes with the velocity that defines the evolution $\vec{v}(\vec{x},t)$ when current density is approximately
\begin{eqnarray}
j_\eta&\equiv& \frac{vB}{\eta} \label{j_eta} \\
&=& \frac{B}{\mu_0 a} R_m.
\end{eqnarray}
The current density needs to be the magnetic Reynolds number $R_m$ times larger than the characteristic current density $B/(\mu_0 a)$.  This can be achieved by the current flowing in thin sheets of width $\sim a/R_m$. 

Written in the form of Equation (\ref{B-adv}), the magnetic field evolution has the mathematical form of an advection-diffusion equation.  Since the 1984 paper by Aref \cite{Aref:1984}, it has been understood that the advection-diffusion equation implies that when the diffusion coefficient $\eta/\mu_0$ becomes extremely small, the timescale for interdiffusive relaxation is approximately $(a/v) \ln(R_m)$.  A well-known example is the temperature of a room reaches a new value of order ten minutes after a radiator is turned on, not after the several weeks as required by thermal diffusion alone.

The fundamental concept that underlies Aref's analysis of the advection-diffusion equation is chaotic trajectories.  There are two parts to the definition of chaotic trajectories.  First, infinitesimally separated trajectories increase their separation  exponentially with distance along the trajectories.  Second, all of the trajectories remain within a bounded distance $a$ of each other across the trajectories.   

Time-dependent divergence-free flows in two dimensions are typically chaotic, $\vec{v}(x,y,t) =\hat{z}\times\vec{\nabla}h(x,y,t)$.  The equations for the trajectories, or streamlines, are $dx/dt=-\partial h/\partial y$ and $dy/dt=\partial h/\partial x$.  A Hamiltonian $h(x,y,t)$ generally has chaotic trajectories when it depends on all three variables but is not when it is independent of a least one.  

Magnetic field lines are defined at a particular point in time and are also given by a Hamiltonian \cite{Boozer:RMP}, $h(\psi,\theta,\varphi)$ with ordinary Cartesian coordines $\vec{x}$ also a function of $(\psi,\theta,\varphi)$.  The implication is magnetic field lines cannot be chaotic unless they have a non-trivial dependence on all three spatial coordinates, but when they do, they are generally chaotic.  Stellarator plasma confinement devices demonstrate that a large volume of non-chaotic field lines is possible even when the magnetic field has a non-trivial dependence on all three spatial coordinates.  Nevertheless, the extreme efforts in design and construction that are required to make stellarator fields non-chaotic demonstrate the unlikelihood of such fields arising naturally.

The two dimensional advection-diffusion equation, 
\begin{equation}
\frac{\partial T(x,y,t)}{\partial t}+\vec{v}\cdot\vec{\nabla}T=D\nabla^2T  \label{T-adv}
\end{equation}
 with $\vec{v}(x,y,t) =\hat{z}\times\vec{\nabla}h(x,y,t)$, relaxes variations in the temperature $T$ in an analogous way to the way the magnetic advection-diffusion Equation (\ref{B-adv}) relaxes tubes that contain magnetic flux \cite{Boozer:2021}.  If $D$ where zero, the  constant-$T$ contours would distort, but not break, and would preserve their enclosed areas.  Highly distorted contours enclosing fixed areas imply temperature gradients become far larger---indeed the maximum gradients become exponentially greater as time advances.  The exponential increase in the gradients implies an arbitrarily small $D$ will relax the temperature variation wherever the gradient is large.  The process continues until temperature gradients disappear on a timescale $\sim (a/v)\ln(va/D)$.   Aref used Lagrangian coordinates to point out the process \cite{Aref:1984}.  In 1999, Tang and Boozer \cite{Tang:1999}  derived the relaxation equation in Lagrangian coordinates and found an effective diffusion coefficient that increases exponentially with time.
 
What will be shown is that the evolution of the magnetic topology, energy, and helicity behave in fundamentally different ways.  Magnetic topology evolves as expected from the advection-diffusion equation $\sim (a/v)\ln(R_m)$.  Magnetic energy can only be dissipated as rapidly as it is added to a system by boundary conditions when $j\sim j_\eta$.  Once magnetic connections can freely break, the magnetic energy goes into Alfv\'en waves, which cause a rapid increase in the current density to $j\sim j_\eta$ that is required for their damping.   Boundary conditions that involve flows with vorticity $\Omega=\hat{b}\cdot\vec{\nabla}\times \vec{v}$, with $\hat{b}\equiv\vec{B}/B$, add helicity to a tube defined by magnetic field lines.  Chaos can spread but cannot destroy magnetic helicity.  The resistive destruction of magnetic helicity depends only on the spatially-averaged current density and cannot be enhanced by concentrating the current density into thin sheets, which is the way energy dissipation is enhanced.  Magnetic helicity added by boundary conditions to a tube defined by magnetic field lines accumulates without limit when $R_m>>1$---at least until the helicity contained in the tube becomes so great that the tube can erupt from the system.

The entanglement of magnetic topology, energy, and helicity evolution in the reconnection literature is sufficiently confusing that sorting out what is important requires a list questions that need to be addressed.  Such a list was provided in 2020 by one hundred eight members of the world reconnection community \cite{Reconnection consensus:2020}, who enumerated nine major challenges to understanding magnetic reconnection.  Their list is copied into Appendix \ref{sec:consensus} and numbered by C1 to C9.  Magnetic reconnection was defined as \emph{the topological rearrangement of magnetic fields}. \emph{Energy conversion} was a challenge, C3.  Magnetic helicity was not specifically mentioned.

The focus of the reconnection community on a current density $j_\eta=vB/\eta$ as a  requirement for reconnection makes acceptance of the fundamental importance of chaos difficult.  This paper is written to be as accessible as possible to anyone who is interested in magnetic reconnection or in addressing the challenges listed by the world reconnection community and copied into Appendix \ref{sec:consensus}. 

\emph{Disentanglement}, Section \ref{disentanglement}, explains how and why magnetic topology, energy, and helicity evolve.  The existing literature on reconnection cannot be placed in context until the distinct aspects of these three types of evolution are understood.

\emph{History of chaos and reconnection}, Section \ref{sec:history}, discusses the literature on reconnection in which all three spatial coordinates enter non-trivially.  In systems with only two non-trivial spatial coordinates, chaos cannot arise and a rapid change in field line connections requires a current density $\sim vB/\eta$ as traditionally expected.  Even in systems that are advertised as fully three-dimensional, the effect of chaos is often omitted or even explicitly excluded with reconnection identified with the need for a current density $\sim vB/\eta$.  A current density $\sim vB/\eta$ is required for rapid Ohmic dissipation of magnetic energy; chaos  provides a uniquely general mechanism for the rapid formation of such a current density.

\emph{Simulations}, Section \ref{simulations}, is on the model developed by Boozer and Elder that is illustrated in Figure \ref{fig:cylinder}.a and studied in a detailed simulation by Huang and Bhattacharjee \cite{Huang-Bhattacharjee}.   The Huang and Bhattacharjee simulation illustrated how magnetic field line connections break when chaos driven by boundary conditions reaches a certain level, independent of the plasma resistivity $\eta$.  After approximately 30\% of the flux available for reconnection had reconnected, the current density quickly rose to a value inversely proportional to $\eta$, which they interpreted as the onset of true reconnection but is also required to balance the power input from the boundary conditions.  Both timescales, the time required for 30\% reconnection and the time for the current density to rise to a value proportional to $1/\eta$, are approximately an order of magnitude longer than the characteristic evolution time, $a/v$ defined by the boundary conditions---independent of $\eta$.  Huang and Bhattacharjee emphasized that the current density reaching a value inversely proportional to $\eta$ showed the irrelevance of chaos to reconnection in solar loops.  Consequently, it is important to understand mathematically what determines the time dependence of the current density.  This is done in \emph{Evolution of the current density},  Appendix \ref{sec:j-evolution}, and \emph{Current density increase of Huang and Bhattacharjee},  Appendix \ref{H-B K-dot}, which are relatively long and more demanding mathematically.

\emph{Runaway electrons and the corona}, Section \ref{corona}, asks whether a current density as intense as $j_\eta=vB/\eta$ in solar loops would have any observational implications.  One such implication could be the corona itself.

\emph{Discussion},  Section \ref{sec:discussion}, points out that although the three aspects of a magnetic evolution topology, energy, and helicity evolve similarly, the topics of primary research interest differ between toroidal plasmas and space and astrophysical plasmas.  Toroidal plasma physicists are focused on magnetic surface breakup while space and astrophysical physicist are focused on energy transfer and may view changing magnetic field line connections of little interest.   However, the change in field line connections through chaos appears be an essential element in a general theory of the rapid transfer of energy.

%%%%%%%%%%%%%%%%%%%%%%%%%%%%%%%%%%%%

\section{Disentanglement \label{disentanglement} } 

The disentanglement of the evolution of magnetic topology, energy, and helicity  is achieved by the mathematical representation of the vector $\vec{E}(\vec{x},t)$ in terms of another vector $\vec{B}(\vec{x},t)$:
\begin{eqnarray}
\vec{E}+\vec{u}_\bot\times\vec{B} &=& - \vec{\nabla}\Phi +\frac{V_\ell}{2\pi}\vec{\nabla}\varphi \mbox{   or  } \label{toroidal}\\
&=&- \vec{\nabla}\Phi +\mathcal{E}\vec{\nabla}\ell\label{non-toroidal}.
\end{eqnarray}
Equation (\ref{toroidal}) was derived by Boozer \cite{Boozer:coordinates} in 1981 for toroidal plasmas with $\varphi$ a toroidal angle and $V_{\ell}$, a constant along magnetic field lines, the loop voltage.  Equation (\ref{non-toroidal}) is a generalization for non-toroidal plasmas with $\mathcal{E}$ constant along magnetic field lines and with $d\ell$ the differential distance along a line.   Both equations are mathematical statements that are fundamentally distinct from Ohm's law, which relates physical quantities such as the plasma mass-flow velocity $\vec{v}$ and the plasma resistivity $\eta$.

In the absence of field nulls, the validity of either Equation (\ref{toroidal}) or (\ref{non-toroidal})  is easily shown.  The component of $\vec{E}$ along $\vec{B}$ is $\vec{E}\cdot\vec{B}= -\vec{B}\cdot\vec{\nabla}\Phi + (V_\ell/2\pi)\vec{B}\cdot\vec{\nabla}\varphi$.  Locally any $\vec{E}\cdot\vec{B}$ can be balanced by $\vec{B}\cdot\vec{\nabla}\Phi$, but for a sufficiently well-behaved $\Phi$ that $\vec{\nabla}\times\vec{\nabla}\Phi=0$ has no subtleties, the potential must be single-valued and finite.  This constraint is imposed by $V_{\ell}$ or $\mathcal{E}$ and may seem enigmatic.  The mathematical importance of $\mathcal{E}$ is clarified by the discussion associated with Equation (\ref{Clebsch-ev}).  Since $\vec{B}$ is non-zero in the region of interest, the term $\vec{u}_\bot\times\vec{B}$ can balance the terms that are perpendicular to the magnetic field.

A magnetic line null can be broken into well separated point nulls by an infinitesimal magnetic perturbation, so only point nulls are physically relevant.  In the presence of point nulls, the derivations of Equations (\ref{toroidal}) and (\ref{non-toroidal}) can be carried though by excluding a small spherical region about each point null.  The potential $\Phi$ at each point null must be chosen so $\oint \vec{j}\cdot d\vec{a}=0$ when integrated over the sphere surrounding the null.  Charge cannot be allowed to accumulate at a point.

%%%%%%%%%%%%%%%%%%%%%%%%%%%

\subsection{Magnetic Topology Evolution}

In 1958, Bill Newcomb \cite{Newcomb} demonstrated that when $\vec{E}+\vec{u}_\bot\times\vec{B} = - \vec{\nabla}\Phi$, magnetic field lines are transported by the flow $\vec{u}_\bot(\vec{x},t)$ and do not change their topology.  The addition of a non-potential term breaks topology.   Following Newcomb, we define the breaking of magnetic field line connections by the existence of a non-zero $V_{\ell}$ or $\mathcal{E}$.  As one would expect, the mass flow velocity of the plasma $\vec{v}(\vec{x},t)$ has no direct effect on the preservation or destruction of magnetic connections---magnetic field line topology preservation is a magnetic and not directly a plasma issue.

The proof that sufficient freedom exists to choose the field lines to be carried by the flow $\vec{u}_\bot$ when and only when $\mathcal{E}=0$ follows from the Clebsch representation, $\vec{B}=\vec{\nabla}\alpha\times\vec{\nabla}\beta$.  Since $\vec{B}\cdot\vec{\nabla}\alpha=0$ and $\vec{B}\cdot\vec{\nabla}\beta=0$, the position of a field line in Cartesian coordinates is $\vec{x}(\alpha,\beta,\ell)$; the Clebsch coordinates are the labels of a field line.  The time derivative
\begin{eqnarray}
\frac{\partial \vec{B}}{\partial t} &=& \vec{\nabla}\times \left( \frac{\partial \alpha}{\partial t}\vec{\nabla}\beta- \frac{\partial \beta}{\partial t}\vec{\nabla}\alpha \right) \\
&=& \vec{\nabla}\times \left( \vec{u}_\bot\times \vec{B} -\vec{\nabla}\Phi -\mathcal{E}\vec{\nabla}\ell\right), \label{Clebsch-ev}
\end{eqnarray}
but $\vec{u}_\bot\times \vec{B}= (\vec{u}_\bot\cdot \vec{\nabla}\beta) \vec{\nabla}\alpha - (\vec{u}_\bot\cdot \vec{\nabla}\alpha) \vec{\nabla}\beta$.  A field line is carried by the flow when its labels $(\alpha,\beta)$ are $\partial\alpha/\partial t+\vec{u}_\bot\cdot \vec{\nabla}\alpha=0$ and $\partial\beta/\partial t+\vec{u}_\bot\cdot \vec{\nabla}\beta=0$.  Sufficient freedom exists for this choice when and only when $\mathcal{E}=0$.

The advection-diffusion equation makes one expect that topology conservation would depend only logarithmically on a small $V_{\ell}$ or $\mathcal{E}$.  The reason is chaos.  Infinitesimally separated magnetic lines increase their separation  exponentially with distance along the lines $\delta(\ell) =\delta_0 e^{\sigma(\ell)}$.  Since all the lines remain within a bounded distance $a$ of each other perpendicular to the magnetic field, the  closest approach of two lines is $\delta_c \approx a e^{-\sigma_{max}}$, where $\sigma_{max}$ is the maximum value of $\sigma(\ell)$ over the distance the lines are followed.  All that is necessary to break the identity of the lines on the scale $a$, is resistive diffusion $\sqrt{(\eta/\mu_0)\tau_{ev}}>\delta_c$ on the evolution timescale of the magnetic field, $\tau_{ev}\equiv a/v$.   In other words, when $\tau_{ev} \approx (\mu_0 a^2/\eta) e^{-2\sigma_{max}}$, the conservation of magnetic field line topology fails, or equivalently when $\sigma_{max}\approx\sqrt{R_m}$.
 
 The concept that magnetic field lines are preserved during an evolution only holds when the distance the lines are followed $\ell$ gives an exponentiation $\sigma(\ell) \lesssim \sqrt{R_m}$.  Magnetic field lines need to be followed only a finite distance to prove magnetic topology is not conserved when it is not.  
 
 Even when $\sigma(\ell) >> \sqrt{R_m}$, boundary conditions on magnetic field lines remain a major problem in space and astrophysics simulations.   The most realistic would appear to be a perfectly conducting, possibly moving, boundary far enough away so only intended interactions arise.  Periodic boundary conditions are frequently used, but they turn the system into a mathematical torus in which field lines that close on themselves can arise on rational surfaces.  As is well known from studies of toroidal plasmas, rational surfaces define the locations of tearing-mode islands, which are plasmoids in space-physics language.  But, the existence of closed field lines seems an implausible as a basis of a theory of magnetic evolution in natural plasmas.  A third possible boundary condition would be an insulating boundary, which means the current density normal to the boundary vanishes.  The magnetic field outside of the boundary is given by the gradient of a potential that is defined by the magnetic field normal to the boundary, which determines the tangential field to the boundary.  Differences in boundary conditions, become important when the magnetic evolution is followed long enough for Alfv\'en waves to travel between the region of interest and the boundary \cite{Boozer:space-rec}.
 
 %%%%%
 
 When $\ln{R}_m>>1$, an important limitation \cite{Boozer:null2019} exists on the ratio of the length $L$ to width the width $a$ of coronal loops for the magnetic field lines to be sufficiently chaotic for the exponentiation to exceed the magnetic Reynolds number:
 \begin{equation} 
 \frac{L}{a} \gtrsim \ln\sqrt{{R}_m}.  \label{L/a cond}
 \end{equation}
 
 The derivation of Equation (\ref{L/a cond}) starts with the equation for the separation $\vec{\delta}$ between infinitesimally separated magnetic field lines $B \partial\vec{\delta}/\partial \ell= (\vec{\delta}\cdot\vec{\nabla})\vec{B}$.  When the tensor $(\vec{\nabla}\vec{B})/B$ is much larger than the scale $a$,  its magnitude is of order $K\equiv \mu_0j_{||}/B$.  Let $d\sigma/d\ell = |\partial\vec{\delta}/\partial \ell|/|\vec{\delta}|$, then the number of e-folds of separation from a given field line $\sigma \approx \int K d\ell\approx KL$.  The change in the magnetic field by the parallel current must be less than the total field, which implies $K\delta_\bot<1$, where $\delta_\bot$ is the width of the current channel.  Consequently, the length of the lines must satisfy $L>\sigma\delta_\bot$.  For chaos to cause a sufficiently rapid changes in field line connections to compete with the evolution, $\delta_\bot$ must be less than $a e^\sigma /\sqrt{R_m}$.  The current channel width, $\delta_\bot$, must be less that the cross-field scale $a$, which implies $\sigma>\ln \sqrt{R_m}$.  The inequality $L/a\gtrsim\ln \sqrt{R_m}$ follows.
 %%%%%
 
 Magnetic topology evolution is the definition of reconnection in the consensus statement of Appendix \ref{sec:consensus}.  It is central to the challenges C1, \emph{The multiscale problem}, C2, \emph{The 3D problem}, C4, \emph{Boundary conditions}, C5, \emph{Onset}, and C6, \emph{Partial ionization}.
 
%%%%%%%%%%%%%%%%%%%%%%% 
 
 \subsection{Magnetic Energy Evolution}
 
 The breaking of the topological constraints on magnetic field lines causes a reduction in the magnetic energy in the large scale magnetic field.  When all topological constraints are removed, the constraint of helicity conservation remains on the energy release.  The helicity constraint is discussed in Section \ref{Helicity evolution}. 
 
 The places where field lines are close together so they can interdiffuse and break their connections are very localized along the magnetic field lines when $R_m>>1$.  This localized breaking leaves field lines out of force balance, and shear Alv\'en waves must propagate along the lines to restore the balance. 
 
 When a sufficiently large amount of magnetic flux has reconnected, large scale force balance is generally lost and the system evolves at an Alfv\'enic rate both along and across the magnetic field lines.  In 2017, Cassk et al \cite{ Cassak:0.1reconnection} reviewed the observation that reconnection commonly proceeds at a rate of 0.1 of the Alfv\'en speed.
 
 The driven shear Alfv\'en waves have two effects.  (1) They tend to concentrate the current into thin sheets as shown in Appendix \ref{sec:j-evolution}.  (2) The concentrated currents  both allow and cause the Alfv\'en waves to be rapidly damped as discussed by a number of authors \cite{Heyvaerts-Priest:1983,Similon:1989,Boozer:flattening}  as well as in Appendix \ref{sec:j-evolution}.  What remains obscure is the ratio of resistive damping, which seems to dominate in the simulations of Huang and Bhattacharjee \cite{Huang-Bhattacharjee}, and viscous damping, which is of equal importance in simple slab models of Alfv\'en wave damping.

 Equation (\ref{non-toroidal}) gives a condition for the transfer of energy between the magnetic field and plasma in a steady-state driven plasma evolution.  The integral of Poynting's Theorem over all of space implies that in steady state $\int \vec{E}\cdot\vec{j} d^3x=0$.  
 
 Equation (\ref{non-toroidal}) implies, 
\begin{equation}
\vec{E}\cdot\vec{j} = \vec{u}_\bot\cdot(\vec{j}\times\vec{B}) + \vec{\nabla}\cdot( \Phi \vec{j})  + (\vec{j}\cdot\vec{\nabla}\ell) \mathcal{E}.
\end{equation}
In a steady state field, the power input to the magnetic field $-\int\vec{u}_\bot\cdot(\vec{j}\times\vec{B})d^3x$ by a moving bounday must be balanced by the power dissipation $\int j_\ell\mathcal{E}d^3x$ with $j_{\ell}\equiv \vec{j}\cdot\vec{\nabla}\ell$.  Assuming $\mathcal{E}$ is produced by $\eta j_{||}$, a steady state requires the current denisty reach an extremely large vale $j_{\ell}\sim u_\bot B/\eta$, which is comparable to $j_{\eta}$ of Equation (\ref{j_eta}).  

Power balance forces the current density along the magnetic field lines $j_{||}$ to reach a level $\sim R_m$ times that required to produce the magnetic field.  A current loop is achieved by having the current flow in a thin sheet of width $\sim a/R_m$ parallel to a field line until is strikes a boundary.  It then flows in the boundary region until it can close its loop by flowing anti-parallel to $\vec{B}$ and reach a boundary on the opposite ends of the magnetic field lines.  See Figure \ref{fig:cylinder}.a and its caption.

 Magnetic energy evolution in the consensus statement of Appendix \ref{sec:consensus} is central to the challenges C1, \emph{The multiscale problem}, C2, \emph{The 3D problem}, C3, \emph{Energy conversion}, and C5, \emph{Onset}.

%%%%%%%%%%%%%%%%%%%%%%%%%%%%%%%%%%%

\subsection{Magnetic Helicity Evolution \label{Helicity evolution} }

In 1958 Woltjer \cite{Woltjer;1958} introduced the concept of magnetic helicity, but it received little notice until its conservation was used by by Bryon Taylor \cite{Taylor:1974} in 1974 to explain remarkable properties of reversed field pinches.  The robustness of magnetic helicity conservation was explained in the 1984 work of Berger \cite{Berger:1984}.  The essential point is that the resistive change in the helicity is given by $\int \eta \vec{j}\cdot\vec{B} d^3x$, but resistive dissipation of energy is given by $\int \eta j^2 d^3x$.  Concentrating the current into a thin channel enhances the energy dissipation but not the rate of helicity change.     As discussed by Boozer and Elder \cite{Boozer-Elder} when the timescale for helicity injection is shorter than the resistive timescale defined by the spatially averaged parallel current, flux-tube eruption must eventually occur.

The evolution of the magnetic helicity is given by $\vec{B}\cdot\vec{E}$, and Equation (\ref{non-toroidal}) implies
\begin{eqnarray}
\vec{B}\cdot\vec{E}&=& -\vec{\nabla}\cdot (\Phi \vec{B})+(\vec{B}\cdot\vec{\nabla}\ell)\mathcal{E} \\
&=& -\vec{\nabla}\cdot (\Phi \vec{B})+B\mathcal{E}.
\end{eqnarray}

 The term $\vec{\nabla}\cdot (\Phi \vec{B})$ integrated over the system volume gives the helicity injection along a magnetic flux tube that penetrates the boundary surrounding the region of interest, $\int \Phi \vec{B}\cdot d\vec{a}$.   After integrating over the angle $\vartheta$ around the flux tube, one can let $\Phi = \Phi_0 +r^2 \Phi''/2+ \cdots$, which gives a flow around the tube $\vec{u}_\bot = (\hat{b}\times\vec{\nabla}\Phi)/B = (r\Phi''/B)\hat{\vartheta}$.  The field line vorticity $\Omega\equiv\hat{b}\cdot\vec{\nabla}\times\vec{u}_\bot=\Phi''/B$ on the enclosing boundary.  The rate helicity is input is then $\Omega (\pi r_0^4/8)B^2$ with $r_0$ the radius of the flux tube.
 
The helicity dissipation is given by $\int B\mathcal{E}d^3x=\int B\eta j_{||} d^3x=\pi r_0^2 L B \eta j_{||}$,
which is essentially proportional to the the volume-averaged current in the plasma---independent of whether it is concentrated into thin sheets or not.   The distance $L$ is the length of the flux tube, and $L_B \equiv \mu_0 j_{||}/2B$ is the distance required for the field line to twist through a radian.

\begin{eqnarray}
\frac{\mbox{helicity input}}{\mbox{helicity dissipation}}&=&\frac{ \Omega (\pi r_0^4/8)B^2}{\pi r_0^2 L B \eta j_{||}} \\ &=&\frac{\Omega r_0^2}{\eta/\mu_0}\frac{L}{16L_B}\sim R_m. \hspace{0.2in}
\end{eqnarray}
The injected magnetic helicity accumulates with negligible resistive destruction for large values of the magnetic Reynolds number.

When footpoint motion injects magnetic helicity into solar loops, there is no systematic way for it to be removed other that by ejecting the entire loop from the sun.

Although helicity is not mentioned in the consensus statement of Appendix \ref{sec:consensus},  helicity evolution is central to the challenges C2, \emph{The 3D problem}, and C9, \emph{Related explosive phenomena}.

%%%%%%%%%%%%%%%%%%%%

\section{History of chaos and reconnection \label{sec:history} }

Once the evolution of magnetic topology, energy, and helicity are disentangled, the reconnection literature can be placed in context. 

One of the best known and highly cited papers on magnetic reconnection in both two and three spatial dimensions is the 1988 article by Schindler, Hesse, and Birn \cite{Schindler:1988} on \emph{General Magnetic Reconnection.}  They focused on magnetic fields that have no points at which $\vec{B}=0$ in the region of interest.   They were aware of the concept of magnetic field line chaos but specifically excluded chaotic fields from consideration.  Consequently, essentially all magnetic fields with a non-trivial dependence on all three spatial coordinates were excluded. With their exclusion of chaos, they found the current density required for reconnection to occur was  approximately equal to $j_\eta$ of Equation (\ref{j_eta}) at the location of the reconnection.  Their definition of magnetic reconnection was based on the tying of the field to the plasma, and magnetic reconnection was not a property purely of the magnetic field.  A section of their paper was titled \emph{Changing Magnetic Helicity}.  They failed to recognize that magnetic helicity is far better conserved than magnetic topology or energy and said: ``global reconnection does in fact imply changing of magnetic helicity."  The Schindler, Hesse, and Birn paper is frequently credited with the commonly-made claim that a current density $\sim j_\eta$ is a requirement for reconnection.  This claim is false for the evolution of the chaotic magnetic fields that were excluded from consideration by Schindler, Hesse, and Birn.

Despite its exclusion by Schindler, Hesse, and Birn, a number of authors have recognized the fundamental importance of chaos.  In 2005, Borgogno, Grasso, Porcelli, Califano,  Pegoraro, and Farina \cite{Borgogno:2005} showed that the interaction of tearing modes with different helicities in toroidal plasmas creates magnetic field chaos and fundamentally changes the definition of magnetic reconnection from the case in which the magnetic field depends on only two spatial coordinates 

In 1999, Lazarian and Vishniac \cite{Lazarian:1999} and in 2011, Eynick, Lazarian, and Vishniac  \cite{Eyink:2011} discussed the role of chaos in the theory of turbulent magnetic reconnection.  This topic was reviewed  by Lazarian, Eyink, Jafari, Kowal, Li, Xu, and Vishniac \cite{Lazarian:2020rev} in 2020.  Turbulent systems are always chaotic, but chaotic systems need not be turbulent.  As explained in the Introduction to  \cite{Boozer:2021}, turbulence slows the effects of chaos as compared to smooth flows with the same speed.  

In 1995 Priest and D\'emoulin \cite{Priest-Demoulin} introduced the concept of \emph{``quasi-separatrix layers,'' where there is a steep gradient in field line linkage}.  Their work recognized the importance of exponentiation.  But, in this and other work, Eric Priest focused on three-dimensional structures that tend to concentrate currents and thereby lead to enhanced breaking of field line connections \cite{Priest:2016} rather than on the absence of a need for concentrated currents when tubes of magnetic field lines become exponentially contorted. 

Reid, Parnell, Hood, and Browning \cite{Reid:2020}, have simulated a case in which the footpoint motions of magnetic field lines do not directly make the lines chaotic but drive large-scale instabilities that do.  

Huang and Bhattacharjee \cite{Huang-Bhattacharjee} recognize that magnetic fields that depend on all three spatial coordinates are generically chaotic and that the chaos makes the maintenance of field line connections fragile.   Despite this, they defined a current density that scales as $1/\eta$ as the ``signature of reconnection."   In their language, the breaking of magnetic connections when the current density is lower is not reconnection---just magnetic diffusion.

The role of helicity conservation in the space sciences is prominent in the theory of dynamos, which is closely related to the theory of reconnection as noted in the consensus document \cite{Reconnection consensus:2020}, Appendix \ref{sec:consensus}.  The conservation properties of the helicity have been known since 1986 to invalidate the $\alpha$-effect dynamo \cite{Boozer:helicity}.  A more detailed proof was given in 1995 by Bhattacharjee and Yuan \cite{Bhattacharjee:1995}.  Nevertheless, the $\alpha$-effect dynamo is commonly studied in dynamo simulations \cite{Rincon:2019} by having a model that destroys helicity at small scales even though this is not energetically possible.  A discussion of helicity conservation in dynamos in the presence of turbulence was given in 2011 by Vishniac and Cho \cite{Vishniac:2001}.  General discussions of helicity transport and dissipation using Lagrangian coordinates are given in \cite{Tang:2000} and \cite{Thiffeault:2003}.
 %%
%%%%%%%%%%%%%%%%%%%%%%
\begin{figure}
\centerline{ \includegraphics[width=3.0in]{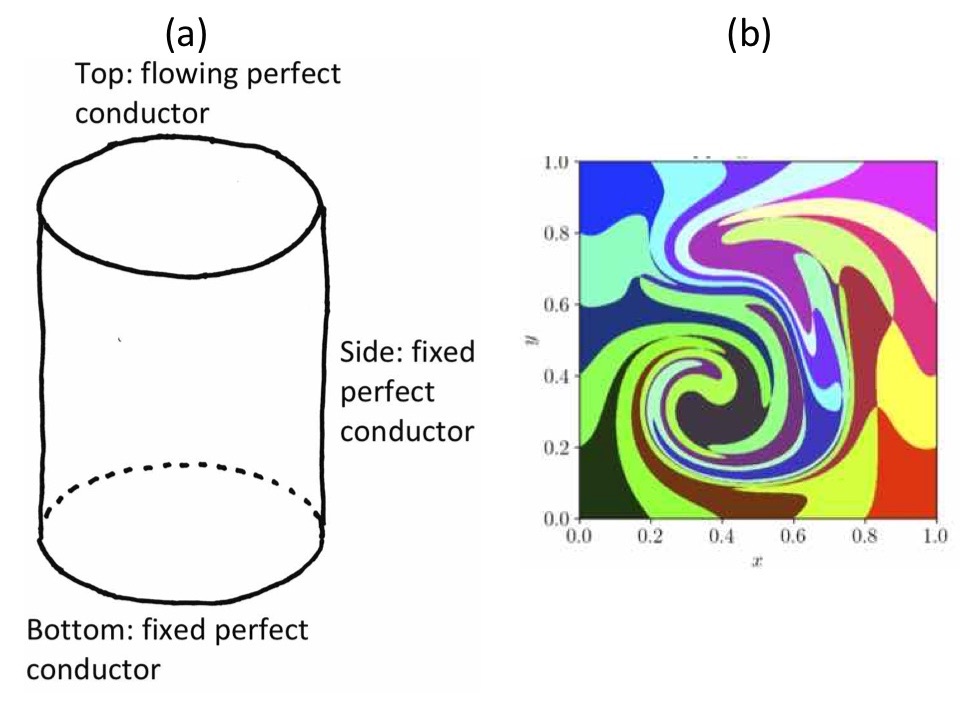}}
\caption{(a) A perfectly conducting cylinder of height $L$ and radius $a$ encloses an ideal pressureless plasma.  All of the sides of the cylinder are fixed except the top, which flows with a specified velocity $\vec{v}_t$.  Initially, $\vec{B}=B_0\hat{z}$.  Each point $\vec{x}_b$ on the bottom of the cylinder defines a line of $\vec{B}$ that in an ideal evolution intercepts a specific point on the top $\vec{x}_t$ with   $\partial \vec{x}_t(\vec{x}_0,t)/\partial t= \vec{v}_t(\vec{x}_t,t)$ and $\vec{x}_0\equiv \vec{x}_t$ at $t=0$. The case of primary interest is when $\vec{v}_t$ is divergence free and chaotic.  This means the $2\times2$ Jacobian matrix  $\partial \vec{x}_t/\partial \vec{x}_0$ has a large singular value that increases exponentially in time and a small singular value that is the inverse of the large singular value.  Reproduced from A. H. Boozer and T. Elder, Phys. Plasmas \textbf{28}, 062303 (2021) with the permission of AIP Publishing.  (b) Huang and Bhattacharjee \cite{Huang-Bhattacharjee} used an equivalent square-cylindrical model to project images on the top boundary of square tubes of magnetic field lines on the bottom boundary.  As the distortions become ever larger, an arbitrarily small resistive diffusion $\eta/\mu_0$ can intermix field lines from different tubes and thereby change their connections.  This figure is part of Figure 5 of their paper. Reproduced from Y.-M. Huang and A. Bhattacharjee,  Phys. Plasmas \textbf{29}, 122902 (2022) with the permission of AIP Publishing. } 
\label{fig:cylinder}
\end{figure}
%%%%%%%%%%%% 

The model defined by Figure \ref{fig:cylinder}.a is closely related to the 1972 Parker Problem \cite{Parker}, which was reviewed by Pontin and Hornig \cite{Pontin-Hornig} in 2020.  Parker thought that footpoint motion would lead naturally to tangential discontinuities---singular current-density sheets---which would give rapid reconnection. What is true it that the breaking of field lines typically occurs at far too low a current density to damp the released energy.  That energy goes first into Alfv\'en waves, which then produce thin sheets of intense current that damp the Alfv\'en waves.  The speed with which this can occur is discussed in Appendix \ref{sec:j-evolution}.  As discussed by Pontin and Hornig, the Parker Problem is often seen as a mechanism for transferring energy from footpoint motion of magnetic field lines into plasma heating.  Although Parker, as did many authors, missed the point that chaos eliminates the need for intense current sheets to break field line connections, the resulting Alfv\'en waves do create such sheets and thereby dissipate the energy released by magnetic-connection changes into plasma thermal energy.

In general, the motions of the footpoints inject magnetic helicity as well as energy.   As discussed in Section \ref{Helicity evolution} and in Boozer and Elder \cite{Boozer-Elder}, helicity dissipation is neither enhanced by chaos nor by intense currents flowing in thin sheets.  When $R_m\rightarrow\infty$, the only way the helicity that is injected into coronal loops can be removed is by the ejection of the loops.  When the magnetic helicity in a plasma region become sufficiently large, force-balance equilibrium is lost, and energy released from the magnetic field goes into the kinetic energy of the large scale plasma flow.

In toroidal plasmas, perturbations can distort the magnetic surfaces and cause rapid large-scale reconnections called disruptions.  But, these perturbations neither inject magnetic helicity nor dissipate significant magnetic energy.  As discussed by Boozer in 2022, the magnetic surfaces can become so contorted \cite{Boozer:surfaces} that the separation between neighboring magnetic surfaces can vary by an exponentially large amount and cause the breaking of surfaces even as $R_m\rightarrow\infty$.  His paper \cite{Boozer:surfaces} motivated a 2022 simulation by Jardin et al \cite{Jardin:2022} that showed the limitation on the electron temperature in a spherical tokamak could be explained by ideal MHD instabilities sufficiently contorting magnetic surfaces to cause breaking despite the smallness of the resistivity.  These results address challenge C4, Boundary conditions, in the list copied in Appendix \ref{sec:consensus}.

%%%%%%%%%%%%

Trying to understand magnetic reconnection when the magnetic Reynolds number is large while ignoring chaos is in defiance of mathematics and intuition.  It is as hopeless as ignoring chaos while trying to understand mixing in fluids when the diffusion coefficient is small.  Nevertheless, the effect chaos on magnetic evolution is subtle because chaos arises only in systems that depend on all three spatial coordinates and has a direct effect only on topological evolution---not on energy or on helicity dissipation. 

The logarithm of the magnetic Reynolds number, $\ln R_m$, is of order of magnitude ten even when it is extremely large.  This leads to a one-sentence statement of the importance of chaos to magnetic reconnection.

\label{sentence}
\begin{quote}
\textbf{Magnetic field lines can go from a simple smooth form to having large and broadly-spread changes in their connections on a timescale that is approximately a factor of ten longer than the ideal evolution time when and only when the magnetic field lines become chaotic.}  
\end{quote}

%\pageref{sentence}

This sentence about chaos could be shown to not be generally valid in two ways: (1) Find an evolving highly chaotic magnetic field that nonetheless preserves well-defined magnetic field line connections.  (2) Find an evolving non-chaotic magnetic field that nonetheless goes from being simple and smooth to large scale connection breaking on a timescale only an order of magnitude longer than the ideal evolution time, even when $R_m$ is many orders of magnitude larger than unity.  

Not all magnetic fields are simple and smooth.  An important example in space physics is a magnetic dipole enclosed in a sphere that interacts with an externally produced magnetic field.  When the dipolar field is sufficiently strong and the external field is simple, two magnetic field nulls are formed outside the dipole sphere \cite{Elder-Boozer:2021}. Then, the preservation of magnetic field line connections has two types of complications that allow connection breaking beyond those in simple smooth fields.  (1) Some field lines go from the dipole sphere to the sources of the external field and information on connection preservation must be transmitted by Alfv\'en waves \cite{Boozer:space-rec}.  (2) The preservation of connections is more difficult when the distance between field lines varies enormously.  Small-scale effects where field lines are close become large-scale effects where the field lines are distant.   As shown by Elder and Boozer \cite{Elder-Boozer:2021} in 2021, field lines that come closer to one of the nulls than $(\delta_s a^2 )^{1/3}$ are indistinguishable, where $\delta_s=c/\omega_{pe}$ or $(\eta/\mu_0)\tau_{ev}$. 

The statement about the importance of chaos can never be proven to be correct.  Karl Popper, one the twentieth century's most influential philosophers of science, famously stated \cite{Popper} that no scientific statement can be proven to be correct but that it must in principle be testable.  The most reliable scientific statements have been tested and never proven false.  

The sentence about chaos has direct implications for five of the nine challenges in the list copied in Appendix \ref{sec:consensus}: C1, \emph{The multiscale problem}; C2, \emph{The 3D problem}; C5, \emph{Onset}; C7, \emph{Flow-driven}; and C9, \emph{Related explosive phenomena}.

%%%%%%%%%%%%%%%%%%%%%%%%%%%%%%%%

\section{Simulations \label{simulations}}

The implications of a given simulation on the various challenges can be subtle.   This is illustrated by a model of a coronal loop driven by footpoint motions, Figure \ref{fig:cylinder}.a from Boozer and Elder \cite{Boozer-Elder}.  This model allows a rigorous monitoring of the topological rearrangement of magnetic fields, and gives an exact expression for the minimum exponentiation of infinitesimally separated field lines in going from one footpoint to the other.

The most complete numerical study of the importance of chaos and current sheets to reconnection was described in a \emph{Featured article} in the Physics of Plasmas, the simulations of Huang and Bhattacharjee \cite{Huang-Bhattacharjee}.  They used the model of Figure \ref{fig:cylinder}.a to study two challenges: C5, \emph{Onset} and C3, \emph{Energy conversion}.   With $R_m=10^4$, they found the breaking of field line connections had reached 30\% of the size of the top surface when the exponentiation of field-line separation approached $10^5$.  Following Boozer and Elder, Huang and Bhattacharjee simplified the interpretation of the simulations by chosing the footpoint velocity $\vec{v}_t$ so no helicity is injected and the ideal, $R_m=\infty$, evolution is stable to kinks.

The reason for large-scale connection breaking when the exponentiation is larger than $R_m$  is intuitively obvious.  Figure \ref{fig:cylinder}.b shows the distortion of tubes of magnetic field lines  that Huang and Bhattacharjee observed for an ideal evolution when he exponentiation was much smaller than $10^5$.  As the distortion of the tubes becomes ever greater, an arbitrarily small resistive diffusivity, $\eta/\mu_0$, can interdiffuse the magnetic fields from different tubes, which changes the field line connections.

As will be shown in Appendix \ref{j-Alfven}, once field-line breaking is large-scale in the model of Figure \ref{fig:cylinder}.a, the current density can quickly rise to a value comparable to $j_{\eta}$ at which resistivity can balance the power input from the footpoint motion.  The Huang-Bhattacharjee  simulations also showed this.  Nevertheless, they interpreted the current density required for power balance as defining the true reconnection.   The onset of large scale topological changes was interpreted as resistive diffusion---not reconnection.  

%%%%%%
When power is continuously put into the plasma, there are two ways it can be dissipated: by resistively $\int\eta j^2d^3x$ and by viscosity $\int\rho\nu (\vec{\nabla}\times\vec{v})^2d^3x$.  An important but unsettled question is what fraction of the power is dissipated by viscosity versus resistivity in the limit as the magnetic Reynolds number $R_m\rightarrow\infty$.  When the viscosity is extremely large, it would appear difficult to for the flow velocity to lie in thin layers, which seems necessary to form the thin layers of intense $j$ that are necessary for a large resistive dissipation.   Is this true only when the fluid Reynolds number $R_f =av/\nu$ is less than unity?  Or, can it be true when the Prandtl number $P_r=R_m/R_f=\mu_0\nu/\eta$ becomes sufficiently large even with $R_f>>1$?  Practical simulations can certainly determine the nature of reconnection for $R_f$ arbitrarily small, but practical limitations on the highest $R_m$ that can be simulated may prevent studies of the case with a very large Prandtl number but with $R_f>1$.  The role of viscosity in turbulent reconnection is addressed analytically in Jafari et al \cite{Jafari:2018}.  The parallel plasma resistivity in plasmas is generally comparable to Spitzer's value for $\eta_{||}$.  However, the ion viscosity can be greatly enhanced by microturbulence \cite{Aschwanden}, the physics of this enhancement is related to the enhancement of the ion thermal conductivity.  Resolution issues also limit simulation tests of viscosity effects in turbulent reconnection.
%%%

When boundary conditions on the magnetic field lines are given at the top and the bottom of the cylinder and $\mathcal{E}=0$ in the plasma, a minimum level of exponentiation in separation is defined independent of the relationship between the plasma velocity $\vec{v}$ and the field line velocity $\vec{u}_\bot$.  This has important implications for challenges C1, \emph{The multiple scale problem}, C2, \emph{The 3D problem}, C4, \emph{Boundary conditions}, and C5, \emph{Onset}, of Appendix \ref{sec:consensus}.

%%%%%%%%%%%%%%%%%%%%
 
Geometric constraints on rapidly reconnection regions are clearly of importance even though they are not  mentioned in the nine fusion challenges.   Simulations could study the way the constraint of Equation (\ref{L/a cond}) manifests itself in the model of Figure \ref{fig:cylinder}.a when the height $L$ to radius $a$ ratio of the cylinder is small compared to $\ln \sqrt{R_m}$.  Do different chaotic regions define separate tubes that undergo reconnection on different time scales?  Boozer and Elder \cite{Boozer-Elder}  found the probability distributions for different levels of exponentiation, which means the fraction of the area of the top in which different levels of exponentiation arose; the median exponentiation is given by the square root of the maximum.  It is not known whether this is a general mathematical result.

%%%%%%%%%%%%%%%%%%%%%%%%%%%%%%%%%%%%%%%%%

 \section{Runaway electrons and the corona \label{corona}}
 
The current density $j_{\eta}$ that is required to Ohmically dissipate the power input of the footpoint motion of solar loops is so large that one might expect observable consequences---a possibility is the existence of the corona itself.

 The current density $j_\eta$ can exceed that required for electron runaway, the Dreicer current density $j_d$. When $j>j_d$, small-angle Coulomb collisions cannot maintain a near Maxwellian distribution, and electrons runaway to a high energy.  The calculations of Kulsrud et al \cite{Kulsrud:1973} imply the rate of electron runaway reaches a significant value at the current density $j_d=2\times10^{-2} e n v_e $.   
 
 As pointed out by Boozer in \cite{Boozer:corona} and discussed  in \cite{Boozer:RMP,Boozer:part-acc}, when the Dreicer current is exceeded, electrons must runaway to whatever energy is required to carry the current.  For the corona, this means to a sufficiently high energy that the electron density $n$ does not become too small due to the gravitational acceleration of the sun $g$.  When the temperature $T$ is constant,  $dn/dr=-n/h$, where the scale height $h\equiv T/Mg$.  When the ionization is high, $M=m_i$, the proton mass, and $h\approx 350T$~km/eV.  A coronal temperature of 100~eV is consistent with a scale height of 35,000 km.  
 
 Below the transition region to the corona, Song \cite{Song:2017} found the electron temperature is almost constant, $\approx0.5$~eV, which implies an electron thermal speed $v_e\approx3\times10^5$~m/s and the Spitzer resistivity $\eta\approx 4\times10^{-3}$~Ohm-meter.  The electron density drops rapidly with altitude above the photosphere and reaches $n\approx3\times 10^{16}/$m$^3$, at the transition.  The Dreicer current at the transition is then $j_d\approx10^5$~A/m$^2$.  The current density $j_{\eta}\equiv vB/\eta\approx 250 vB$, which equals $j_d$ when $Bv\approx400~$T$\cdot$m/s.  Song found the magnetic field is highly localized in flux tubes on the photosphere, but those tubes have large diameters at the transition region.
 
A $vB$ product for estimating $j_{\eta}$ can be obtained for coronal loops driven by sunspots.  Okamoto and Sakurai \cite{Sunspot B} have observed fields above 0.6~T at sunspots, and   Sobotka and Puschmann \cite{Solar flow} have observed horizontal flows of 4~km/s.  The product of these numbers gives $Bv=2400$~T$\cdot$m/s approximately six times higher than that required for $j_d=j_{\eta}$ at the transition.  More typical velocities and fields could produce an exact balance.
 
 As noted in  \cite{Boozer:RMP}, any star that has evolving magnetic field structures on the scale of tens of thousands of kilometers must have a corona, otherwise the induced currents would run out of current carriers, but whether this actually explains the solar corona requires careful study.  
 
 Although coronal heating is not directly mentioned among the challenges of Appendix \ref{sec:consensus}, it is connected with C3, \emph{Energy conversion}, and C6, \emph{Partial ionization}.

 %%%%%%%%%%%%%%%%%%%%
 
 %%%%%%%%%%%%%%%%%%%%%%%%%%%%%%%%%%%%%%

\vspace{0.2in}

\section{Discussion \label{sec:discussion}}

Disentangling the evolution of magnetic topology, energy, and helicity is required for a sensible discussion of the scientific challenges associated with reconnection \cite{Reconnection consensus:2020}, which is quoted in Appendix \ref{sec:consensus}.  Even so, the important questions differ between studies of toroidal magnetic-fusion plasmas and studies of space and astrophysical plasmas: (1) When the initial condition of an evolving magnetic field is smooth, is the time required for reconnection to occur on a timescale comparable to the timescale set by an evolution?  This is a central issue in tokamak disruptions, but the onset time for reconnection has not been a focus in space and astrophysical plasmas.  (2) How should the speed of reconnection be defined?  When rapid energy transfer from the magnetic field to the plasmas is the definition of reconnection then the rate of transfer provides a definition.  In toroidal plasmas, the time scale for disruption effects to arise is more important.

\subsection{Reconnection in toroidal plasmas}

The periodicity of toroidal fusion plasmas gives a clear definition of the breaking of magnetic connections.  Breaking connections means breaking magnetic surfaces, as in a tokamak disruption.

When a rapid breaking of the toroidal magnetic surfaces occurs in a tokamak, the definition of the speed of reconnection is subtle.  Magnetic field lines are defined at points in time.  When the last intact magnetic surface is broken, a magnetic field line at that instant changes from being bound by that surface to traversing the plasma and striking the chamber walls.  The relevant speed is not defined by the instantaneous change in the trajectory of the field line but by the speed of physical effects that are produced by the topological change.  This is the time it takes for particles or energy to be transported along magnetic field lines throughout a chaotic region.  For $j_{||}/B$ flattening, the characteristic time is the time for shear Alfv\'en waves to cover the chaotic region by propagating along the magnetic field lines.  The topological change can allow relativistic electrons trapped in the core of a tokamak to strike the surrounding walls by following magnetic field lines.  The damage to the device is largely determined by how highly localized in space and time are the strikes of the relativistic electrons on the walls, which is determined by the speed with which the topology of the magnetic field lines change \cite{Boozer-Punjabi:2016}.

%%%%%%%%%%%%%%%

\subsection{Reconnection in space and astrophysical plasmas}

In space plasmas the boundary conditions are often too indeterminate to rigorously define magnetic field line topology or what is meant by the breaking of field-line connections.  Consequently, little study is done of the physical effects produced by the changes in field line connections.  Yet, answers to physical questions within the region of interest may depend on what happens outside that region \cite{Boozer:space-rec}.

When topology and changes in field line connections are ill-defined, energy transfers between the field and the plasma seem most important, and it is natural to define magnetic reconnection by the energy transfer \cite{Hesse-Cassak}.  Energy transfer can be defined even in models in which changes in magnetic topology are not defined.  Energy release from the magnetic field can occur even when $\eta=0$; ideal magnetic kinks are a well known example.  Nevertheless, the energy release from the magnetic field is generally greater when the field-line connections are freely broken.  The direct energy release from the magnetic field  is of little interest in the fast reconnections called tokamak disruptions for less than a part in a thousand of the energy is typically released \cite{Boozer:2017}.

Space and astrophysical studies are focused not only on the energy transfer from the fields to the plasma but also on acceleration of particles by the reconnection process.  Although the acceleration of electrons to relativistic energies as a result of magnetic surface reconnection is a major issue in tokamaks, the acceleration is not a direct part of the reconnection process but rather a result of the plasma cooling increasing the resistivity to the point that $\eta \bar{j}$, with $\bar{j}$ a spatially-averaged current density, gives an electric field above that required for electron runaway.  Electron runaway also offers an explanation of the solar corona and serves as a check on the production of intense currents by even smooth, large-scale footpoint motion.  

%%%%%%%%%%%

\subsection{Commonality of reconnection issues}

The centrality of chaos to understanding changes in the topology of magnetic field lines seems clear when the evolution involves all three spatial coordinates.  It is implied by a mathematical analysis of Faraday's law, by intuition based on pictures such as Figure \ref{fig:cylinder}.b, and by analogy to mixing in fluids.

By making the preservation of connections of magnetic field lines fragile, chaos can cause energy to be transferred from the large scale magnetic field to Alfv\'en waves.  The damping of the Alfv\'en  waves involves intense current sheets.  The extent to which the fluid viscosity of the plasma can dissipate the energy released by the magnetic field requires more study.  At a sufficiently large Prandtl number, $P_r\equiv \nu/(\eta/\mu_0)$, the ratio of viscosity to resistivity, the flows needed to produce the intense current sheets must be damped, but the level is presently unknown.

Neither magnetic field line chaos nor intense current sheets enhance the dissipation of magnetic helicity. Helicity input from footpoint boundary conditions cannot be dissipated in a low resistivity plasma, which makes helicity accumulation an obvious cause for eruption of coronal loops.  Helicity conservation during the magnetic reconnection of tokamak disruptions, limits the release of energy from the magnetic field to extremely small values \cite{Boozer:2017}. 

\subsection{Importance of chaos in the reconnection literature}

In 1962, Thomas Kuhn \cite{Kuhn} in \emph{one of the most influential works of history and philosophy written in the 20th century} \cite{Britannica:Kuhn},  discussed how differences in the important questions and in the important features of a model arise whenever a new paradigm is introduced.  He also pointed out how difficult it is for a scientific community to accept a change in paradigm.  

Although the one sentence statement about chaos on page \pageref{sentence} should be tested, the fact that Faraday's law and the general expression for the electric field give an advection-diffusion equation make failures unlikely in the extreme.  Basing a reconnection theory on ignoring mathematics or Maxwell's equations does not seem a winning strategy.  A change in the paradigm for rapid changes in magnetic topology is required from intense current sheets with $j\approx vB/\eta$ of the Schindler, Hesse, and Birn analysis, in which chaos is explicitly ignored, to the centrality of chaos.

Simulations  of magnetic evolution with credible boundary conditions as the largest achievable magnetic Reynolds numbers are needed to guide the physics.  Many issues could be better understood through practical simulations.  A few topics of importance to solar physics are (1)  the constraint $L/a\gtrsim \ln \sqrt{R_m}$ on coronal loops, (2) the role of viscosity, (3) the extent to which Dreicer runaway explains the corona, (4) the constraints of solar footpoint flows that can be obtained from the behavior of coronal loops, (5) the effect of field-line curvature on the evolution of coronal loops.  

Unfortunately, simulations do not have the numerical resolution to directly solve magnetic evolution problems with the largest magnetic Reynolds numbers of practical interest.  These problems require an analytic understanding of evolution of the three physical concepts: magnetic topology, energy, and helicity.

%%%%%%%%%%%%%%%%

%%%%%%%%%%%%%%%%%%%%%%%%%%%%%%%%%%%%%%%%%%%%%%%%%%%%
 
\section*{Acknowledgements}
This work was supported by the U.S. Department of Energy, Office of Science, Office of Fusion Energy Sciences under Award Numbers DE-FG02-95ER54333, DE-FG02-03ER54696, DE-SC0018424, and DE-SC0019479.

\section*{Data availability statement}

Data sharing is not applicable to this article as no new data were created or analyzed in this study.

%%%%%%%%%%%%%%%%%%%%%%%%%%%%%%%%%%%%%%%%%%%%%%%%%%%%

%%%%%%%%%%%%%%%%%%%%%%%%%%%%%%%%%%%%%%%%%%%%%%%%%%
%%%%%%%%%%%%%%%%%%%%%%%%%%%%%%%%%%%%%%%%

\appendix

%%%%%%%%%%%%%%%%%%%%%%%%%%%%%%

\section{\label{sec:consensus} }

\textbf{Major Scientific Challenges and Opportunities in Understanding Magnetic Reconnection and Related Explosive Phenomena in Solar and Heliospheric Plasmas \cite{Reconnection consensus:2020}}\\

\textbf{I. Magnetic Reconnection: A Fundamental Process throughout the Universe and in the Lab}\\

Magnetic reconnection---the topological rearrangement of magnetic fields---underlies many explosive phenomena across a wide range of natural and laboratory plasmas. It plays a pivotal role in electron and ion heating, particle acceleration to high energies, energy transport, and self-organization. Reconnection can have a complex relationship with turbulence at both large and small scales, leading to various effects which are only beginning to be understood. In heliophysics, magnetic reconnection plays a key role in solar flares, coronal mass ejections and heating, the interaction of the solar wind with planetary magnetospheres, associated dynamical phenomena such as magnetic substorms, and the behavior of the heliospheric boundary with the interstellar medium. Magnetic reconnection is also integral to the solar and planetary dynamo processes. In short, magnetic reconnection plays a key role in many energetic phenomena throughout the Universe, including extreme space weather events that have significant societal impact and laboratory fusion plasmas intended to generate carbon-free energy.\\

\textbf{II. Major Scientific Challenges in Understanding Reconnection and Related Explosive Phenomena
in Heliophysics}\\

\textbf{C1. The multiple scale problem:} Reconnection involves the coupling between the fluid or MHD scale of the system and the kinetic ion and electron dissipation scales that are orders of magnitude smaller. This coupling is currently not well understood, and the lack of proper treatments in a self-consistent model is the core of the problem. Reconnection phase diagrams based on plasmoid dynamics clarify different possibilities for coupling mechanisms. Key questions include: how do plasmoid dynamics scale with key parameters, such as the Lundquist number and effective size; how is this scaling influenced by a guide field; do other coupling mechanisms exist; and how does reconnection respond to turbulence and associated dissipation on scales below or above the electron scales? \\

\textbf{C2. The 3D problem:} Numerous studies have focused on reconnection in 2D while natural plasmas are 3D. It is critical to understand which features of 2D systems carry over to 3D, and which are fundamentally altered. Effects that require topological analysis include instabilities due to variations in the third direction leading to complex interacting ``flux ropes," potentially enhancing magnetic stochasticity, and field-line separation in 3D. How fast reconnection is rlated to self-organization phenomena such as Taylor relaxation, as well as the accumulation of magnetic helicity, remains a longstanding problem with important implications for, e.g., coronal heating and eruptions. \\

\textbf{C3. Energy conversion:} Reconnection is invoked to explain the observed conversion of magnetic energy to heat, flow, and to non-thermal particle energy. A major challenge in connecting theories and experiments to observations is the ability to quantify the detailed energy conversion and partitioning processes. Competing theories of particle acceleration based on 2D and 3D reconnection have been proposed, but as of yet there is no consensus on the origin of the observed power laws in particle energy distributions.\\

\textbf{C4. Boundary conditions:} It is unclear whether an understanding of reconnection physics in periodic systems can be directly applied to natural plasmas, which are non-periodic and often line-tied at their ends such as in solar flares. Whether line-tying and driving from the boundaries fundamentally alter reconnection physics has profound importance in connecting laboratory physics to heliophysics. It is also important to learn how reconnection works in naturally occurring settings that have background flows, out-of-plane magnetic fields, and asymmetries.\\

\textbf{C5. Onset:} Reconnection in heliophysical and laboratory plasmas often occurs impulsively, with slow energy build up followed by a rapid energy release. Is the onset a local, spontaneous (e.g., plasmoid instability) or a globally driven process (e.g., ideal MHD instabilities), and is the onset mechanism a 2D or 3D phenomenon? How do collisionality and global magnetic geometry affect the onset conditions? A related question is how magnetic energy is accumulated and stored prior to onset, e.g., in filament channels on the Sun and in the lobes of Earth's magnetotail.\\ 

\textbf{C6. Partial ionization:} Reconnection events often occur in weakly ionized plasmas, such as the solar chromosphere (whose heating requirements dwarf those of corona), introducing new physics from neutral particles. Questions include whether reconnection is slowed by increased friction or accelerated by enhanced two-fluid effects.\\ 

\textbf{C7. Flow-driven:} Magnetic fields are generated by dynamos in flow-driven systems such as stars and planets, and reconnection is an integral part of the dynamo process. Key questions include: under what conditions can reconnection occur in such systems; how fast does it proceed; how does reconnection affect the associated turbulence? \\

\textbf{C8. Turbulence, shocks, and reconnection:} Reconnection is closely interconnected to other fundamental plasma processes such as turbulence and shocks, which in turn produce heliophysical phenomena such as solar energetic particles. It is essential to understand the rates of topology change, energy release, and heating during reconnection, as they may be tied to the overall turbulence and shock dynamics.\\

\textbf{C9. Related explosive phenomena:} Global ideal MHD instabilities, both linear (kink, torus) and nonlinear (ballooning), are closely related to reconnection either as a driver or a consequence (e.g., coronal mass ejections, magnetic storms/substorms, and dipolarization fronts in the magnetotail). Understanding how, and under what conditions, such explosive phenomena take place, as well as their impact, remain major scientific challenges. Physics insights from reconnection under extreme conditions in astrophysics should be beneficial as well.

%%%%%%%%%%%%%%%%%%%%%%%%%%%%

%%%%%%%%%%%%%%%%%%

\section{Evolution of the current density \label{sec:j-evolution}}

The increase in the current density along an arbitrarily chosen magnetic field line and the increase in the current density in a given flow are the two ways to estimate the rate the plasma current increases.

Surprisingly, the formation of current sheets and current densities comparable to $j_{\eta}$ of Equation (\ref{j_eta}) is not mentioned in the list of challenges in Appendix \ref{sec:consensus}.  Nevertheless, it enters many discussions about reconnection.

%%%%%%%%%%%%

\subsection{Current density along an arbitrary line of $\vec{B}$ \label{sec:j-B} }

\subsubsection{The near-line expansion}

A Taylor expansion near an arbitrarily chosen magnetic field line $\vec{x}_0(\ell,t)$, called the central line, gives the relation between the change in the twist of line with the distance along the line $\ell$ and the time derivative of the current density $j_{||}$ along the line \cite{Boozer:2022}.

The derivation uses the position vector $\vec{x}(\rho,\alpha,\ell)=\rho \cos\alpha\hat{\kappa}_0 + \rho\sin\alpha \hat{\tau}_0+\vec{x}_0(\ell,t)$, where $\rho$ is the distance from the central line; $\hat{\kappa_0}$ and $\hat{\tau}_0$ are the curvature and torsion unit vectors of the cental line $\vec{x}_0(\ell,t)$.  The trajectories of the adjacent lines are given by a  $\tilde{H}=\tilde{\psi}h(\alpha,s)$, where  $h=k_\omega(s,t) + k_q(s,t)\cos\big(2\alpha-\varphi_q(s,t)\big)$ with $k_\omega\equiv K_0/2 + \tau_0$. The magnitude of the quadrupole component of the magnetic field is given by $k_q\tilde{\psi}$, the only $\rho^2$ order Fourier component in the Hamiltonian.  The magnetic flux is $\tilde{\psi}\equiv \pi B_0 \rho^2$.

When the ideal evolution term, $\vec{u}_\bot\times\vec{B}$, is large compared to the resistive term, the evolution of the parallel current $j_{||}$ along an arbitrarily chosen magnetic field line $\vec{x}_0(\ell,t)$ is given by \cite{Boozer:2022,Boozer:2022E}
\begin{equation}
\frac{\partial \Omega B_0}{\partial \ell} =\frac{\partial\left(K_0 + 2\tau_0 +\frac{4k_q^2}{\partial\varphi_q/\partial s}\right)B_0}{\partial t}. \label{corr}
\end{equation}
$K_0(\ell,t)\equiv \mu_0j_{||}/B_0$, $\Omega(\ell,t)=\hat{b}_0\cdot \vec{\nabla}\times\vec{u}_\bot$, and $\tau_0(\ell,t)$ is the torsion of the curve.  The quantity $k_q^2/(\partial\varphi_q/\partial\ell)$ is defined by the quadrupole contribution to the Hamiltonian for the adjacent magnetic field lines and should only be retained when the adjacent field lines are not chaotic.

A more intuitive and more easily interpreted form of Equation (\ref{corr}) is
\begin{equation}
\frac{\partial \Omega B_0}{\partial \ell} =\frac{\partial\left(K_0 + 2\nu\right)B_0}{\partial t}, \label{nu-K}
\end{equation}
where $\nu(\ell,t)$ is the stellarator-like rotational transform per unit length of field lines  produced by currents at a distance from the chosen line.  The term on the right-hand side of Equation (\ref{nu-K}) comes from the fact that when the externally-produced rotational transform $\nu$ per unit length has a positive time derivative, then $K_0$ must decrease to keep the total transform per unit length fixed.  The factor of two comes from dependence of the current-produced transform at a radius $\rho_0$ being proportional to $(\int K \rho d\rho)/\rho_0^2=K_0/2$ as $\rho_0\rightarrow0$.  The left-hand side comes from the fact that when the field lines are undergoing a twist per unit time, which is what $\Omega$ is, then a current must increase to produce a total rotational transform per unit length that gives  $\partial\Omega/\partial\ell$.

The term $\nu$ in Equation (\ref{nu-K}) represents the effects of distant currents.  In principle, distant currents can produce chaos in a bounded region by themselves.  For example, the magnetic surfaces in a curl-free stellarator can be perturbed to produce chaotic regions using additional curl-free fields that resonate with the rational surfaces.   But, the term $\nu$ also represents the natural response of a perfectly conducting medium to a changing current.  Le Chatelier's principle states that nature tends to resist change.  One would expect a larger $K_0$ would be required to produce a given field line twist due to the presence of distant currents.  Indeed, this is what was seen in the simulations of Huang and Bhattacharjee \cite{Huang-Bhattacharjee} as is discussed in Appendix \ref{H-B K-dot}.

%%%%%%%%%

\subsubsection{Force-balance and plasma flow}

Once magnetic field lines have become chaotic, so connections are easily broken, the power input goes into plasma motion until it can be dissipated by resistivity or viscosity.

The divergence-free nature of the current $\vec{j}$ implies $K$ and the Lorentz force $\vec{f}_L\equiv\vec{j}\times\vec{B}$ are related by
\begin{eqnarray}
\frac{\partial K}{\partial\ell} &=& \hat{b}\cdot\vec{\nabla}\times\frac{\mu_0 \vec{f}_L}{B^2},   \mbox{   where   }\\
\vec{f}_L&=& \rho \left( \frac{\partial \vec{v}}{\partial t}+ \vec{v}\cdot\vec{\nabla}\vec{v} -\nu \vec{\nabla}\times\vec{\Omega}\right). \label{Navier-Stokes}
\end{eqnarray}
When the linear inertial term dominates $V_A^2\partial K/\partial\ell=\partial \Omega/\partial t$ and $K$ relaxes to being uniform along magnetic field lines at the Alfv\'en speed, $V_A$.

%As discussed in Section \ref{sec:HB}, the behavior of $\partial K/\partial t$ depends on the relative magnitudes of $\partial\vec{v}/\partial t$, $\vec{v}\cdot\vec{\nabla}\vec{v}$, and $\nu \vec{\nabla}\times\vec{\Omega}$, which could be explored by simulations.

In principle, the distinction between the velocity of the plasma $\vec{v}$ and that associated with the magnetic field $\vec{u}$ could complicate the force-balance equation, but this distinction not generally not considered important in a near-ideal plasma.

%%%%%%%%%%%%

\subsubsection{Increase in $j$ due to Alfv\'en waves \label{j-Alfven} }

The current density required for breaking magnetic connections scales as $\ln R_m$ while the current density required to balance the power input into coronal loops by footpoint motion scales as $R_m$. The power released by the breaking of field line connections must initially go into Alfv\'en waves when $R_m$ is large and viscosity effects are small, as in the simulations of Huang and Bhattacharjee \cite{Huang-Bhattacharjee}.  Consequently, the energy in Alv\'en waves increases until the current density reaches the value $j_\eta$, Equation (\ref{j_eta}), at which resistive dissipation can balance the power input of the footpoint motion.  Alfv\'en energy is evenly divided between magnetic and kinetic energy with the velocity and magnetic field fluctuations related by  $\tilde{v}/V_A=\tilde{B}/B$ with $V_A$ the Alfv\'en speed.

The increase in the current density to the level the energy and Alfv\'en waves can be rapidly damped in essentially the topic of Alfv\'en wave damping in chaotic magnetic fields, which has been studied by several authors \cite{Heyvaerts-Priest:1983,Similon:1989,Boozer:flattening}.

The field-line localized formula for the current density $\partial K/\partial t=\partial\Omega/\partial \ell$, where $K\equiv \mu_0j/B_0$, predicts a rapid increase in the current density to the level $j_\eta$, or equivalently $K_\eta\equiv \mu_0j_\eta/B_0\approx R_m/L$, required to balance the input power due to the contribution of the Alfv\'en waves to the vorticity $\Omega$.

Once the velocity fluctuations $\vec{\tilde{v}}$ of the Alfv\'en waves are large, it is natural to expect $|\partial\vec{v}/\partial t| \approx| \vec{v}\cdot\vec{\nabla} \vec{v}| \approx \tilde{v}^2/\delta_{tan}$ where $\delta_{tan}$ is the scale of $\vec{\tilde{v}}$ variation along itself.  This scale is much longer than the distance scale $\delta_\bot$ for $\vec{\tilde{v}}$ changes across the flow.  This disparity in scales $\delta_{tan}>>\delta_\bot$ is related to the concentrated current lying in thin but broad ribbons along the magnetic field lines as seen by Boozer and Elder and predicted in \cite{Boozer:2021} and illustrated in Figure 6 of Huang and Bhattacharjee's paper \cite{Huang-Bhattacharjee}.  Since Alfv\'en waves propagate with the Alfv\'en speed, $|\partial\tilde{v}/\partial t| \approx (V_A/\delta_{\ell})\tilde{v}$, with $\delta_\ell$ the spatial scale parallel to B.  Consequently, $\delta_{\ell}\approx (V_A/\tilde{v})\delta_{tan}$.

The contribution of Alfv\'en waves to the vorticity is $\Omega_A\approx \tilde{v}/\delta_\bot$, so $\partial\Omega_A/\partial \ell \approx \Omega_A/\delta_{\ell}$.  The fraction of the cross sectional area occupied by the Alfv\'en waves and the sheets of intense current density is $f_A=\delta_\bot\delta_{tan}/a^2$, so  \begin{eqnarray}
\frac{\partial\Omega_A}{\partial \ell} &\approx& \left(\frac{\tilde{v}}{V_A}\right)^2 \frac{V_A}{f_A a^2} \\
&\approx& \left(\frac{\tilde{B}}{B}\right)^2 \frac{V_A}{f_A a^2}.
\end{eqnarray}

The time required to set up the sheet current
\begin{eqnarray}
K_s&=&\frac{\mu_0}{\eta}\frac{v_ta}{L}=\frac{R_m}{L} \hspace{0.1in}\mbox{is   }\\
\tau_s &=& \frac{K_s}{\partial\Omega_A/\partial \ell} \\
&\approx& \left(\frac{B}{\tilde{B}}\right)^2 \frac{f_A a^2}{V_A} \frac{R_m}{L}\\
&\approx& \left(\frac{aB}{L\tilde{B}}\right)^2\frac{L}{V_A}
\end{eqnarray}
since $f_AR_m\approx1$.  %In order to track the footpoints $L\tilde{B}/aB\approx1$ as well.

The Introduction to \cite{Boozer:null2019} explains why the magnetic perturbation perpendicular to the initial magnetic field reaches an amplitude $\Delta B\approx (a/L) \ln R_m$ when large scale reconnection commences.  The argument is essentially that of Equation (\ref{L/a cond}).  A large fraction of the associated magnetic energy goes into the Alfv\'en waves so time to build up the sheet current can be very short $\tau_s \approx (L/V_A)/(\ln R_m)^2$ compared to the time required to reach sufficient exponentiation $(a/v_t)\ln R_m$.  The velocity of the footpoints $v_t$ is small compared to the Alfv\'en speed $V_A$ in problems of primary interest.  Once large scale breaking of field line connections had occurred, the formation of sheet currents quickly followed in the simulations of Huang and Bhattacharjee \cite{Huang-Bhattacharjee}.

%%%%%%%%%%

\subsection{Current increase in a given flow}

When the magnetic field lines have a known velocity, $\vec{u}_\bot(\vec{x},t)$, the Cauchy solution for the ideal evolution of the magnetic field is
\begin{eqnarray}
\vec{B}\big(\vec{x}(\vec{x}_0,t)\big)=\frac{\tensor{J}_L}{\mathcal{J}_L}\cdot\vec{B}(\vec{x}_0), \label{Cauchy}
\end{eqnarray}
where $\tensor{J}_L$ is the Jacobian matrix of the Lagrangian coordinates of $\vec{u}_\bot$ and $\mathcal{J}_L$ is the determinant of $\tensor{J}_L$.  The history of this solution was reviewed by Stern \cite{Stern:66} in 1966.

Equation (\ref{Cauchy}) has profound implications about the differences in  between two and three dimensional evolution and the speed with which the current density can increase.  

The Cauchy solution is a purely mathematical statement about Faraday's law, $\partial\vec{B}/\partial t=-\vec{\nabla}\times\vec{E}$, and the representation of the vector $\vec{E}$ in terms of $\vec{B}$ with $\mathcal{E}=0$, Equation (\ref{non-toroidal}).

Appendix \ref{Lag-coord} defines Lagrangian coordinates, $\vec{x}(\vec{x}_0,t)$, and explains the Singular Value Decomposition of the Jacobian matrix $\tensor{J}_L\equiv\partial\vec{x}/\partial\vec{x}_0$.  The implications of the Cauchy solution are discussed in Appendix \ref{Cauchy-imp}.

\subsubsection{Lagrangian coordinates \label{Lag-coord} }

Lagrangian coordinates $\vec{x}_0$ are defined so that the position vector in ordinary Cartesian coordinates is $\vec{x}(\vec{x}_0,t)$, where
\begin{equation}
\left(\frac{\partial \vec{x}}{\partial t}\right)_L  \equiv \vec{u}_\bot(\vec{x},t)   \mbox{    with   } \vec{x}(\vec{x}_0,t=0)=\vec{x}_0.
\end{equation}
The subscript ``$L$" implies the Lagrangian coordinates $\vec{x}_0$ are held fixed.

The three-by-three Jacobian matrix of Lagrangian coordinates can be decomposed as
\begin{eqnarray}
\frac{\partial \vec{x}}{\partial \vec{x}_0} &\equiv& \left(\begin{array}{ccc}\frac{\partial x}{\partial x_0} & \frac{\partial x}{\partial y_0} & \frac{\partial x}{\partial z_0} \vspace{0.03in}  \\ \vspace{0.03in} \frac{\partial y}{\partial x_0} & \frac{\partial y}{\partial y_0} & \frac{\partial y}{\partial z_0}  \\ \frac{\partial z}{\partial x_0} & \frac{\partial z}{\partial y_0} & \frac{\partial z}{\partial z_0}\end{array}\right) \nonumber\\
  &=& \tensor{U}\cdot\left(\begin{array}{ccc}\Lambda_u & 0 & 0 \\0 & \Lambda_m & 0 \\0 & 0 & \Lambda_s\end{array}\right)\cdot\tensor{V}^\dag . \label{SVD.of.Jacobian}
\end{eqnarray}
where $\tensor{U}$ and $\tensor{V}$ are unitary matrices, $\tensor{U}\cdot\tensor{U}^\dag=\tensor{1}$.   The three real coefficients $\Lambda_u \geq \Lambda_m \geq \Lambda_s \geq 0$ are the singular values of the Singular Value Decomposition (SVD).  The Jacobian matrix can also be written as \
\begin{equation}
\frac{\partial\vec{x}}{\partial \vec{x}_0} = \hat{U}\Lambda_u \hat{u} + \hat{M}\Lambda_m \hat{m} + \hat{S}\Lambda_s \hat{s},
\end{equation}
where $\hat{U}$, $\hat{M}$, and $\hat{S}$ are orthogonal unit vectors, $\hat{U}=\hat{M}\times\hat{S}$, of the unitary matrix $\tensor{U}$, which means they define directions in the ordinary space of Cartesian coordinates, $\vec{x}$.  The unit vectors $\hat{u}$, $\hat{m}$, and $\hat{s}$ are determined by the unitary matrix $\tensor{V}$, which means that they define directions in the space of Lagrangian coordinates, $\vec{x}_0$.

The Jacobian of Lagrangian coordinates, which is the determinant of the Jacobian matrix, is 
\begin{eqnarray}
\mathcal{J}_L&=&\Lambda_u \Lambda_m \Lambda_s.
\end{eqnarray}
The time derivative $\big(\partial \ln(J_L)/\partial t\big)_L = \vec{\nabla}\cdot\vec{u}_\bot$.  For the model of Figure \ref{fig:cylinder}.a, the Jacobian changes little from unity.

The properties of evolving magnetic fields and currents using Lagrangian coordinates were discussed by Tang and Boozer \cite{Tang:2000} in 2000 and by Thiffeault and Boozer \cite{Thiffeault:2003} in 2003.

%%%%%%%%%%%%%

\subsubsection{Implications of the Cauchy $\vec{B}(\vec{x},t)$  \label{Cauchy-imp}}

Using the Singular Value Decomposition of Lagrangian coordinates, Equation (\ref{Cauchy}) for the Cauchy solution implies \cite{Boozer:2021} 
\begin{eqnarray}
B^2 = \left( \frac{\hat{u}^\dag\cdot\vec{B}_0}{\Lambda_m\Lambda_s}\right)^2 +  \left( \frac{\hat{m}^\dag\cdot\vec{B}_0}{\Lambda_u\Lambda_s}\right)^2+ \left( \frac{\hat{s}^\dag\cdot\vec{B}_0}{\Lambda_u\Lambda_m}\right)^2.\hspace{0.06in}
\end{eqnarray}

The mathematical definition of a chaotic $\vec{u}_\bot$ is that the largest singular value $\Lambda_u >  \exp(t/\tau_L)$ for some $\tau_L>0$ for any time $t$ greater than a sufficiently large value.  The smallest $\tau_L$ that satisfies this inequality gives the exponentiation timescale.    The product of the three singular values $\Lambda_u \Lambda_m \Lambda_s\approx1$.  When $\Lambda_u$ increases exponentially, $\Lambda_s$ decreases exponentially, and $\Lambda_m$ has at most an algebraic dependence on time.  The exponentiation time scale $\tau_L$ is usually comparable to the evolution time scale.

The term in $B^2 $ proportional to $(\hat{u}^\dag\cdot\vec{B}_0)^2$ goes to infinity exponentially in time.   The term  proportional to $(\hat{s}^\dag\cdot\vec{B}_0)^2$ goes to zero exponentially.   A bounded magnetic field strength is only possible for a time long compared to $\tau_L$ when the magnetic field points in the $\hat{M}$ direction, 
\begin{equation}
\vec{B}(\vec{x},t) \rightarrow \frac{\hat{m}^\dag\cdot\vec{B}_0}{\Lambda_u\Lambda_s}\hat{M}.
\end{equation}
The unit vector $\hat{M}$ is also the unit vector along the magnetic field $\hat{b}$. When the magnetic field is in the $\hat{M}$ direction the current density $\vec{j}$  lies in ribbons along the magnetic field lines which become exponentially wider and exponentially thinner in time with the magnitude of the current density increasing only algebraically \cite{Boozer:2021}.  The results of Boozer and Elder \cite{Boozer-Elder} in 2021 exhibit these properties as apparently does the ideal solution of Huang and Bhattacharjee \cite{Huang-Bhattacharjee}.

Any smooth flow must naturally be consistent with the magnetic field lying in the $\hat{M}$ direction.   Otherwise the magnetic field pressure $B^2/2\mu_0$ would increase exponentially in time.  

The number of spatial dimensions is critical in reconnection theory because the number of singular values of Jacobian matrix equals the number of coordinates.  In two dimensions, a chaotic $\vec{u}_\bot$ implies the magnetic field strength must increase exponentially in time, but not in three dimensions.

When $\vec{B}$ has a small component in the $\hat{U}$ direction, that component and the associated current density are amplified exponentially in time until that component becomes comparable to $\vec{B}$.  The implication is that localized flows, which are seen in the Huang and Bhattacharjee simulations \cite{Huang-Bhattacharjee}, can produce thin current sheets on a fast time scale as  discussed in Appendix \ref{j-Alfven}.  

Although force-limits  push $\vec{B}$ to be in the direction $\hat{M}$, which allows only an algebraic increase in current density with time, the small deviations in $\vec{B}$ that are associated with different current profiles can be in the exponentiating $\hat{U}$ direction, which has an exponentially increasing current density. This exponentiation can only hold while the current channel narrows while the spatially-averaged current density remains essentially constant.

%%%%%%%%%%%%%%%%%%%%%%%%%%%%%%% 
\section{Current density increase of Huang and Bhattacharjee \label{H-B K-dot}}

Equation (15) of Huang and Bhattacharjee \cite{Huang-Bhattacharjee} appears to be analogous to Equation (\ref{nu-K}).   Their analogue to $2\partial\nu(\ell,t)/\partial t$ is $\mathcal{T}$, where
\begin{equation}
\mathcal{T}\equiv \partial_x\vec{u} \cdot \partial_y \vec{B}_\bot - \partial_y\vec{u}\cdot\partial_x \vec{B}_\bot. \label{T-math}
\end{equation}
Their paper emphasizes the importance of $\mathcal{T}$ to the differences between their results and those of Boozer and Elder \cite{Boozer-Elder}.

Near a given line the divergence-free magnetic field line velocity and the ideal perturbation to the magnetic field $\delta\vec{B}$ can be written in terms of the field line displacement $\vec{\Delta}$ with $\vec{\nabla}\cdot\vec{\Delta}=0$;
\begin{eqnarray}
\vec{u} &=& \frac{\partial \vec{\Delta}}{\partial t};\\
\delta\vec{B}&=&\vec{\nabla}\times(\vec{\Delta}\times B_0\hat{z})\\
&=&B_0\frac{\partial\vec{\Delta}}{\partial z}.
\end{eqnarray}
The displacement is the sum of two parts: a part with a curl, $\vec{\Delta}_c$, and a quadrupole part, $\Delta_q$, that does not.  Each has an associated velocity and perturbed magnetic field.  The displacement $\Delta_c$ also has an associated vorticity $\Omega$ and a parallel current density, or $K$, but $\Delta_q$ does not.
\begin{eqnarray}
\vec{\Delta}_c &=& \Delta_{c1}(z,t) x\hat{y} - \Delta_{c2}(z,t) y\hat{x};\\
\Omega &\equiv&\hat{z}\cdot \vec{\nabla}\times\vec{u}_c \\
&=& \frac{\partial( \Delta_{c1}+ \Delta_{c2})}{\partial t};\\
\frac{\delta\vec{B}_c}{B_0} &=& \frac{\partial \Delta_{c1}}{\partial z} x\hat{y} - \frac{\partial \Delta_{c2}}{\partial z} y\hat{x}; \\
K&\equiv& \frac{\hat{z}\cdot\vec{\nabla}\times\delta\vec{B}_c}{B_0}\\
&=&  \frac{\partial( \Delta_{c1}+ \Delta_{c2})}{\partial z};\\
\vec{\Delta}_q &=&\Delta_q(t)(x\hat{x}-y\hat{y})\cos(k_zz) \nonumber\\
&&+ \Delta_q(t)(y\hat{x}+x\hat{y})\sin(k_zz);\\
\delta\vec{B}_q&=&k_zB_0\Big\{(y\hat{x}+x\hat{y})\cos(k_zz)\nonumber\\
&&-(x\hat{x}-y\hat{y})\sin(k_zz)\Big\}.
\end{eqnarray}
The magnetic scalar potential that gives the dipolar field is proportional to $\rho^2 \cos(2\theta-k_zz)$, where $\rho$ is the distance from the line, and $\theta$ is the angle around the line.  Only one Fourier component in $z$ is retained, but an arbitrary number of $k_z$'s could be.  The second harmonic is the only curl-free term that can contribute in $\rho^2$ order.

The there are four contributions to $\mathcal{T}$ of Equation (\ref{T-math}).  $\mathcal{T}=\mathcal{T}_{cc} + \mathcal{T}_{cq}+\mathcal{T}_{qc}+\mathcal{T}_{qq}$ with first suffix denoting $\vec{u}_c$ or $\vec{u}_q$ and the second the corresponding $\delta\vec{B}$.
\begin{eqnarray}
\mathcal{T}_{cc}&=&0;\\
\mathcal{T}_{cq}&=&k_zB_0 \frac{\partial( \Delta_{c1}- \Delta_{c2})}{\partial t} \Delta_q \sin(k_zz);\\
\mathcal{T}_{cq}&=&B_0 \frac{\partial \Delta_q}{\partial t} \frac{\partial( \Delta_{c1}- \Delta_{c2})}{\partial z}\cos(k_zz);\\
\mathcal{T}_{qq}&=&k_zB_0\frac{d\Delta_q^2}{dt}.
\end{eqnarray}
Letting $\Delta_{c1}- \Delta_{c2} =\Delta_c(t)\cos(k_zz) + \Delta_s(t)\sin(k_zz)$ gives
\begin{eqnarray}
\mathcal{T} %&=&k_zB_0 \Big\{ \left(\frac{d\Delta_c}{dt} \Delta_q-\Delta_{c}\frac{d\Delta_q}{dt}\right)\frac{\sin(2k_zz)}{2} \nonumber\\ &&+\frac{d\Delta_s}{dt} \Delta_q\sin^2(kz) + \Delta_{s}\frac{d\Delta_q}{dt} \cos^2(k_zz)\nonumber\\&&+\frac{d\Delta_q^2}{dt}\Big\}\\
&=& k_zB_0 \left(\frac{d(\Delta_q^2+ \Delta_s\Delta_q/2)}{dt}\right) \nonumber\\&&+ k_zB_0\Big\{ \left(\frac{d\Delta_c}{dt} \Delta_q-\Delta_{c}\frac{d\Delta_q}{dt}\right)\frac{\sin(2k_zz)}{2} \nonumber\\&&+ \left(\Delta_{s}\frac{d\Delta_q}{dt} -\frac{d\Delta_s}{dt} \Delta_q\right)\frac{\cos(2k_zz)}{2}\Big\}.
\end{eqnarray}

When the various $\Delta$'s increase together, so $d\ln\Delta_c/dt=d\ln\Delta_s/dt=d\ln\Delta_q/dt$, 
\begin{eqnarray}
\mathcal{T}&=&k_zB_0\left(\frac{d(\Delta_q^2+ \Delta_s\Delta_q/2)}{dt}\right) \hspace{0.1in}\mbox{and}\hspace{0.1in}\\
\nu_{hb}&=&k_z\frac{\Delta_q^2+ \Delta_s\Delta_q/2}{2a^2}
\end{eqnarray}
is the Huang and Bhattacharjee analogue of $\nu$ in Equation (\ref{nu-K}) using the coefficients that they used to make their equations dimensionless.

%%%%%%%%%%%%%%%%%%%%%%%%%%%%%%%%%%%%%%%%%%%%%%%%%%%%%%%%%%%%%%%%%%%%%%%%%%%%%%%%%%%%%%%%%%%%%%%%%%%%%%%%%%

\end{document}